\documentclass[9pt,twocolumn,twoside]{opticajnl}
\journal{opticajournal} % use for journal or Optica Open submissions

\usepackage{amsmath,amssymb}  
\usepackage{multirow}
\setboolean{shortarticle}{true}
\usepackage{lineno}
\title{Embedding Matrices in Programmable Photonic Networks with Flexible Depth and Width}

\author[1,2]{Matthew Markowitz}
\author[2]{Kevin Zelaya}
\author[1,2,*]{Mohammad-Ali Miri}

\affil[1]{Department of Physics, Queens College of the City University of New York, Queens, New York 11367, USA}
\affil[2]{Physics Program, The Graduate Center of the City University of New York, New York, New York 10016, USA}
\affil[*]{Corresponding author: mmiri@qc.cuny.edu}

\begin{abstract}
\noindent
We show that programmable photonic circuit architectures composed of alternating mixing layers and active layers offer a high degree of flexibility. This alternating configuration enables the systematic tailoring of both the network's depth (number of layers) and width (size of each layer) without compromising computational capabilities. From a mathematical perspective, our approach can be viewed as embedding an arbitrary target matrix into a higher-dimensional matrix, which can then be represented with fewer layers and larger active elements. We derive a general relation for the width and depth of a network that guarantees representing all $N \times N$ complex matrix operations. Remarkably, we show that just two such active layers—interleaved with passive mixing layers—are sufficient to universally implement arbitrary matrix transformations. This result promises a more adaptable and scalable route to photonic matrix processors.

\end{abstract}

\setboolean{displaycopyright}{false} % Do not include copyright or licensing information in submission.

\begin{document}

\maketitle

%\section{Introduction}

Photonic integrated circuits (PIC) are an important platform in next-generation computing, enabling near-light-speed information processing for resource-intensive tasks in telecommunications, artificial intelligence, and quantum computing~\cite{perez2017multipurpose,Siew:21,zhang2021optical,fu2024optical}. Programmable matrix-vector multiplication (MVM) technologies outstand the race towards wide-scale PIC manufacturing~\cite{carolan2015universal,bogaertsProgrammablePhotonicCircuits2020,zhou2022photonic}, given their ubiquity as a fundamental operator in computing. A plethora of research has been conducted to develop compact, low-power, and error-resilient $N$-port architectures capable of approximating arbitrary $N \times N$ matrices. Particularly, meshes of Mach–Zehnder interferometers (MZI) have been shown to cover all elements of the unitary group $U(N)$~\cite{Reck94}, leading to the development of integrated photonic units that perform arbitrary unitaries through triangular~\cite{miller2013self,de2018simple}, rectangular~\cite{Clements16}, and other mesh topologies~\cite{Shokraneh2020,rahbardar2023addressing}. Alternatively, a lower-depth MZI mesh has allowed the reduction of active elements to about half for a subclass of dense unitaries~\cite{tang2024lower}. Parallel to these developments, an alternative class of circuit architectures based on alternating phase shifter arrays with fixed mixing layers has been explored \cite{tangIntegratedReconfigurableUnitary2017a, saygin_robust_2020, pastorArbitraryOpticalWave2021, markowitz2023universal, zelaya2024goldilocks, Markowitz23Auto}. In general, the minimum number of phase layers has been demonstrated via numerical analysis to be $M=N+1$, valid for a large class of mixing layers with proper dense transfer matrices~\cite{markowitz2023universal,zelaya2024goldilocks}, which contributes to the overall robustness and error-resilience of the design~\cite{Markowitz23Auto}.

\begin{figure*}[t]
\centering %\flushleft
\includegraphics[width=0.85\textwidth]{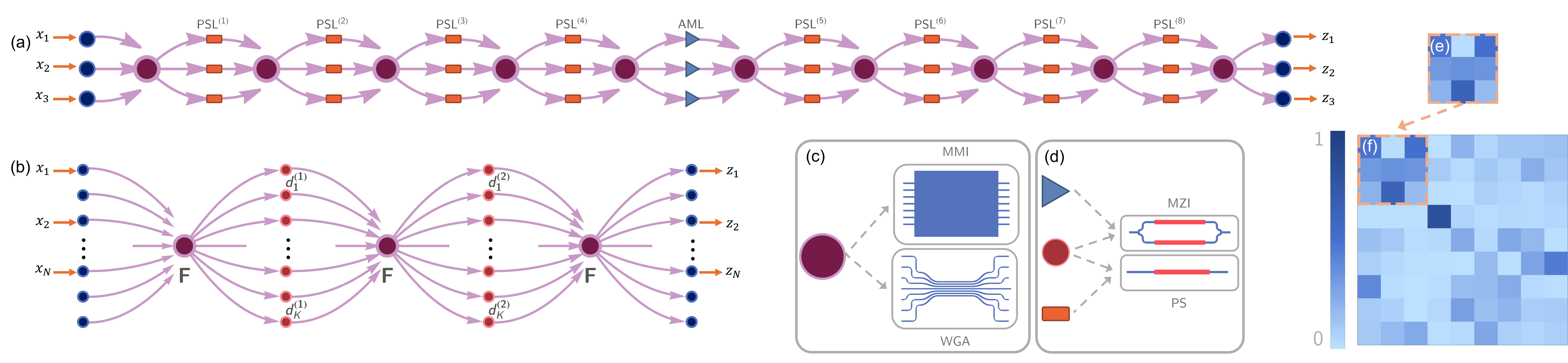}
\caption{Operational graph of the conventional SVD architecture (a) and proposed flexible two-layer architecture (b). The mixing layers $F$ alternate with diagonal phase layers with elements $d_{p}^{(m)}$, with $p\in\{1,\ldots,K\}$ and $m\in\{1,2\}$. $F$ can be realized using MMI couplers or coupled waveguide arrays (c). The active diagonal layer can comprise lossy elements, such as MZIs or PCMs, or lossless phase shifters (d). The intensity of the desired $N \times N$ target matrix (e) is encoded into the two-layer $K \times K$ system (f). 
}
\label{fig1}
\end{figure*}

Research surrounding the mesh architectures and the alternating architectures has predominantly concentrated on achieving universality within the $U(N)$ framework as a fundamental transformation domain in optical processing. In turn, constructing arbitrary complex-valued matrices involves employing singular value decomposition (SVD) \cite{miller2012all,miller2013self,giamougiannisUniversalLinearOptics2023} (Fig.~\ref{fig1}(a)). %This technique facilitates the decomposition of general matrices into a product of two unitary matrices and a positive semidefinite diagonal matrix \cite{Moon00}, where the diagonal matrix can be realized through layers of amplitude modulators, such as phase-change material (PCM)~\cite{Rios2022ultra}, lossy waveguides~\cite{dave2024clearing}, or conventional MZIs~\cite{miller2013self}. See Fig.~\ref{fig1}(a) . 
The extension to non-unitary matrix operations has been found useful in orthogonal communication channels~\cite{miller2000communicating,miller2019waves} and direction-of-arrival finding sensors~\cite{zelaya2024photonic}. Recently, we have shown that general complex-valued matrices can be realized directly through the alternating architectures with $N+1$ programmable layers that involve simultaneous amplitude and phase modulation~\cite{markowitz2024learning}. 
%\textcolor{blue}{However, a core limitation of all existing architectures is their rigidity. Their fixed topology restricts the ability to optimize critical device parameters such as geometry, insertion loss, or bandwidth.} 
Therefore, it is of great interest to find alternative structures that implement complex-valued matrices directly and with flexible designs that allow, for example, a smaller transverse or longitudinal footprint.

This letter introduces a matrix-vector multiplication (MVM) architecture based on a truncated alternating design, which minimizes the number of layers and effective active ports used in the MVM operation. This approach optimizes the use and placement of the active elements, enabling the reconstruction of target matrices embedded within higher-dimensional matrices. An exact relation between the number of layers, total number of ports, and active elements required to reconstruct arbitrary complex-valued matrices is derived, which is further numerically tested. This approach ultimately allows for efficiently reshaping the MVM architecture by scaling it horizontally or vertically, according to the design needs.

%\textcolor{blue}{- Here, we show exact formula for tuning the width and depth }
%By encoding an $N\times N$ matrix into a $2N\times 2N$ port system, it is shown that most $N+3$ phase layers are required to approximate $N\times N$ complex matrices, effectively reducing horizontal footprint by increasing that of the vertical \cite{tang2024lower}.

%\section{Methods}
The proposed architecture is introduced as a low-footprint device that performs general linear transformations. This is achieved through the $K$-port and $M$-layered interlaced device illustrated in Fig.~\ref{fig1}(b), with $K\in\mathbb{Z}^{+}$ and restricted to $M\in\{2,3\}$ layers. Although arbitrary $K\times K$ matrices cannot be reconstructed in such a low-depth device, as reported in previous works~\cite{Markowitz23Auto,zelaya2024goldilocks,markowitz2024learning}, one can aim for reconstructing smaller-dimensional $N\times N$ matrices ($N<M$). This process is henceforth called matrix-embedding, and it is illustrated in the insets of Fig.~\ref{fig1}. The interlaced architecture intertwines passive mixing layers $F\in U(K)$ with complex-valued diagonal matrices $D\in\mathbb{C}^{K\times K}$, where $D^{(m)}_{p,q}=d^{(m)}_{p}e^{i\phi^{(m)}_{p}}\delta_{p,q}$ and $0\leq d^{(m)}<1$. The diagonal layer incorporates both phase and amplitude modulation to increase the available degrees of freedom and learning capabilities of the architecture. This choice induces a wave evolution of the form
\begin{equation}
%A = F D^{(M)} F D^{(M-1)} F D^{(M-2)} \cdots D^{(3)} F D^{(2)} F D^{(1)} F,
A^{(2)} = F D^{(2)} F D^{(1)} F, \quad 
A^{(3)} = F D^{(3)} F D^{(2)} F D^{(1)} F, 
\label{univ_A}
\end{equation}
where, $A^{(M)}\in\mathbb{C}^{K\times K}$ for $M\in\{2,3\}$ interlacing layers composed by the product $D^{(m)}F$. An additional mixing layer $F$ has been included at the end of the architecture, as it is found to improve the optimization results. 

From~\eqref{univ_A}, two cases are worth discussing: the \textit{unitary design} where all the parameters are restricted to phase elements ($d^{(m)}_{p}=1$), and the \textit{non-unitary design} that incorporates additional losses through amplitude modulators ($d^{(m)}_{p}\neq 1$). See Fig.~\ref{fig1} for details. Since both designs allow controlled power leakage through some unused (dummy) ports, both designs are intrinsically lossy. The mixing layers $F$ are taken as either the DFT or discrete fractional Fourier transform (DFrFT). See Figs.~\ref{fig1}(c)-(d). However, the mixing layer is not tied to a specific choice of $F$ as long as its transmission matrix fulfills the required density property~\cite{zelaya2024goldilocks}.

Generally, an arbitrary complex-valued matrix $S\in\mathbb{C}^{N\times N}$ requires $2N^2$ parameters for its representation. If $S$ is embedded in $A$, a minimum bound for $K$ is found by estimating the number of parameters in $A$ that fully- or over-parameterizes $S$. The straightforward calculations lead to the relations $2KM-2M+2 \geq 2N^2-1$ and $MK-M+1\geq 2N^2-1$ for the non-unitary and unitary designs, respectively. A phase has been factored out out of $M-1$ layers, which induces a $U(1)$ gauge transformation~\cite{Markowitz23Auto}, yielding to the critical values
\begin{equation}
K_{c}^{(U)} = \left\lfloor\frac{2N^2-2}{M}+1 \right\rfloor , \quad K_{c}^{(nU)} = \left\lfloor\frac{2N^2-3}{2M}+1 \right\rfloor , 
\label{K_crit}
\end{equation}
for the unitary and non-unitary designs, respectively. By convention, the embedding is taken to be in the upper-left corner of the larger matrix when using the DFT as the mixing layer (Figs.~\ref{fig1}(e)-(f)), whereas different embeddings are used for the DFrFT case. The amplitude modulation thus serves as a means by which the stage number may be reduced by up to a factor of two. Furthermore, because the lossy diagonal elements can be implemented by replacing each phase shifter with an MZI switch, the horizontal footprint increases only slightly due to the inclusion of the arms. We therefore investigate how the width and height of the circuit, parameterized by $M$ and $K$ for a given $N$, can be modified while still exhibiting universality.

For every $N$, $K$ and $M$, 100 complex-valued random matrices $S$ are generated using the SVD, $ S \equiv U \Sigma V^{\dagger}$,
%\begin{equation}
%    S = U \Sigma V^{\dagger}
%\end{equation}
where $U$ and $V$ are $N \times N$ unitary matrices sampled from the Haar distribution, and $\Sigma$ is a positive semidefinite diagonal matrix with entries uniformly chosen from the interval $[0,1]$. Since the distribution of the singular values $\Sigma$ depends explicitly on the matrix under test, all targets are normalized through $\widetilde{S}:=\sqrt{N} S/Tr(SS^{\dagger})$ so that $Tr(S^{\dagger}S) = N$. The loss function for the optimization becomes
\begin{equation}
\label{eq_L}
L = \| \alpha \widetilde{A} - \widetilde{S} \|^2/N^{2} ,
\end{equation}
where $\widetilde{A}\in\mathbb{C}^{N\times N}$ is the corresponding submatrix of $A\in\mathbb{C}^{K\times K}$ and $\alpha^{-2}$ is a scaling factor representing the total loss ratio. Universality is only met once a certain threshold value $\alpha_{crit}$ is surpassed. The case $\alpha > \alpha_{crit}$ shows traces of universality as the loss elements are re-scaled. For each optimization, $\alpha$ is held fixed, and so to probe universality for various parameter choices, $\alpha$ is initially set to a large value ($\alpha \approx 10$). The optimization is performed using the gradient-based method we previously implemented in~\cite{Markowitz23Auto,zelaya2024goldilocks,markowitz2024learning}, with 100 runs per target until the heuristically determined threshold $L_c = 10^{-7}$ is met. 

%\clearpage 
\begin{figure}[ht]
\centering %\flushleft
\includegraphics[width=0.75\linewidth]{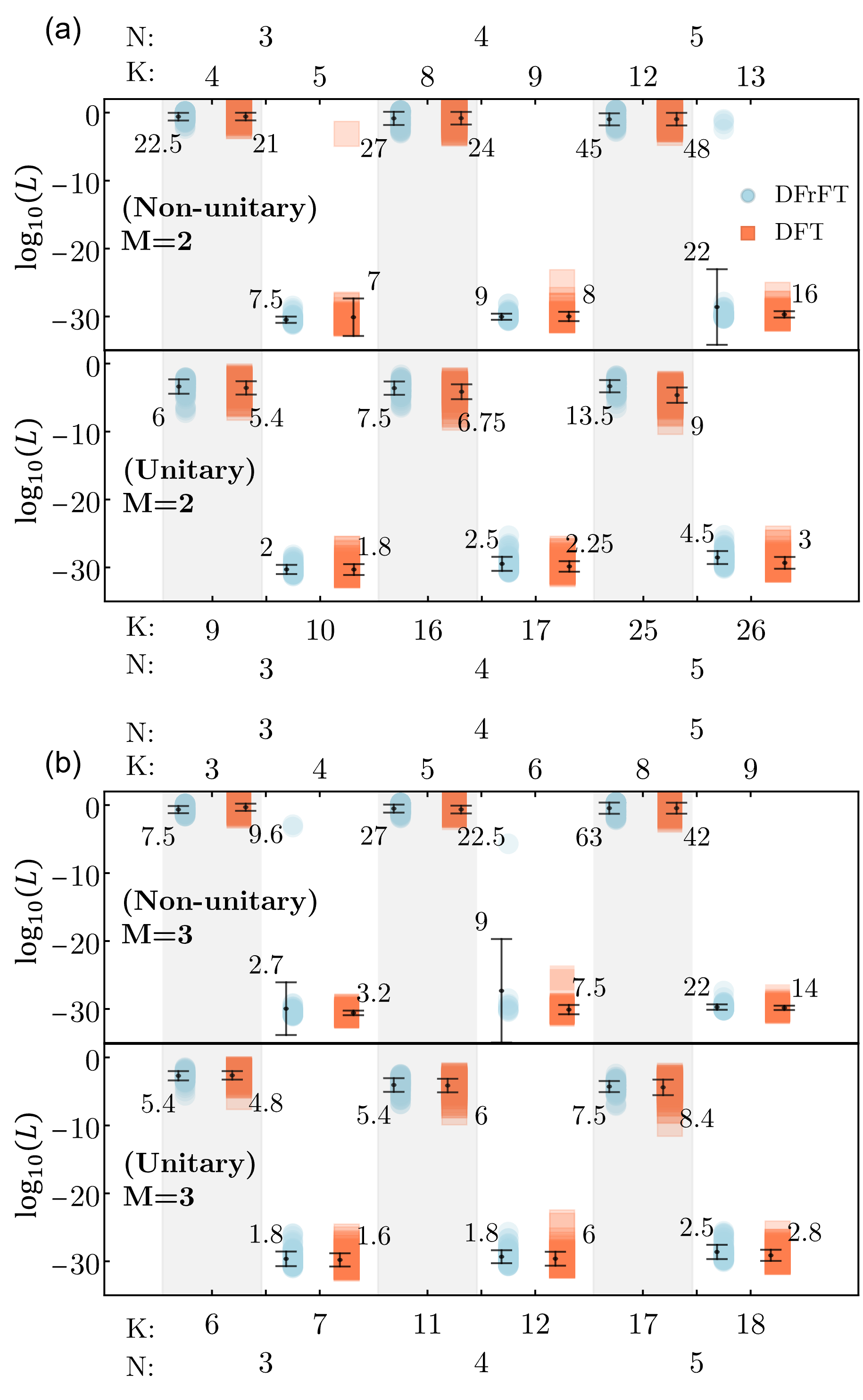}
\caption{Error norm $\textnormal{log}_{10}(L)$ for 100 random targets and embedded matrices of size $N=3,4,5$ and various choices of $K$. The number of layers is fixed to $M=2$ (a) and $M=3$ (b). In each case, the upper and lower panels show the error norms for the non-unitary ($\vert d^{(m)}_{p}\vert\neq 1$) and unitary ($\vert d^{(m)}_{p}\vert= 1$) designs, respectively. Complex-valued $\mathbb{C}^{N\times N}$ targets are considered during the optimization, combined with the DFT and DFrFT as the mixing layers $F$. The number next to the error bars denotes the corresponding power-loss factor $\alpha$.}
\label{fig2}
\end{figure}

Different shapes of the embedding matrix $\widetilde{A}$ are investigated for the cases with and without amplitude modulation for $N\in\{3,4,5,6\}$. For each $M\in\{2,3\}$, the loss function is given for embedding sizes $K\in\{ K_{c},K_{c}+1 \}$. As shown in Fig.~\ref{fig2}, the loss drops by several orders of magnitude between $K=K_{C}$ and $K=K_{c}+1$, the former having values typically found in the literature for under-parameterized and "bad" architectures, which suggests that $K=K_{C}+1$ is indeed required to enter the regime of universality.

%\section{Results}
\begin{figure}[ht]
\centering %\flushleft
\includegraphics[width=0.45\textwidth]{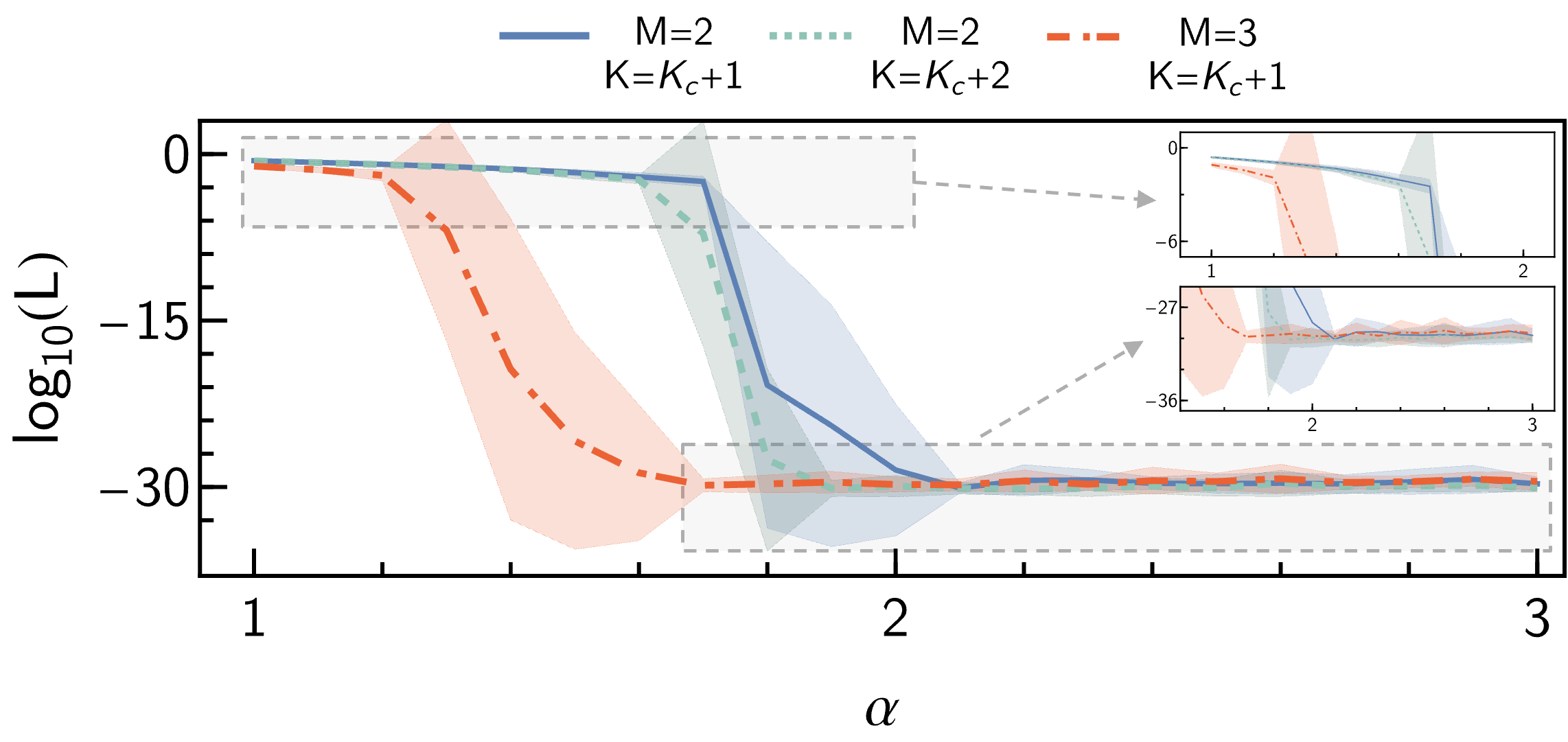}
\caption{Loss-function (log$_{10}(L)$) dependence on the scaling parameter $\alpha$ for the targets in $\mathbb{C}^{4\times 4}$. The unitary architecture is considered, with the DFT as the passive $F$ layer. Only values of $K>K_{c}$ above the critical value are considered for $M=2$ and $M=3$ layers. For the analysis, 50 targets are chosen per $\alpha\in[1,3]$, from which the mean (curve) and standard deviation (shaded area) are displayed.}
\label{fig3}
\end{figure}

Since the overall goal is to reduce the horizontal length of the circuit, the case $M=2$ is of particular interest. In Fig.~\ref{fig3}, the loss function~\eqref{eq_L} is plotted as a function of $\alpha$ for $N=4$ with $K\in\{K_{C}+1,K_{C}+1\}$, each for multiple target matrices. These reveal the $\alpha_{crit}$, approximately 1.8 for $M=2$ and 1.5 for $N=3$ at $K=K_{C}+1$, which are then chosen as the $\alpha$ parameter in the optimization to determine the optimal values $d_p^{(m)}$ and $\phi_p^{(m)}$. As these critical points increase with $N$, a fundamental issue arises for large $N$ where the loss may become too large. As seen in Fig.~\ref{fig3}, this can be combated, perhaps surprisingly, by increasing either $M$ or $K$. Nonetheless, the large number of loss elements for large circuits becomes a non-trivial issue. Adding optical amplifiers in some layers can potentially mitigate deleterious effects, however, at the cost of increased width and complexity. Theoretically, if amplifiers are located in each layer, the entire device can be re-scaled so that $\alpha = 1$ and the maximum amplitude gain can be taken as $\alpha_{crit}^{1/M}$. 

\begin{figure*}
    \centering
    \includegraphics[width=0.8\linewidth]{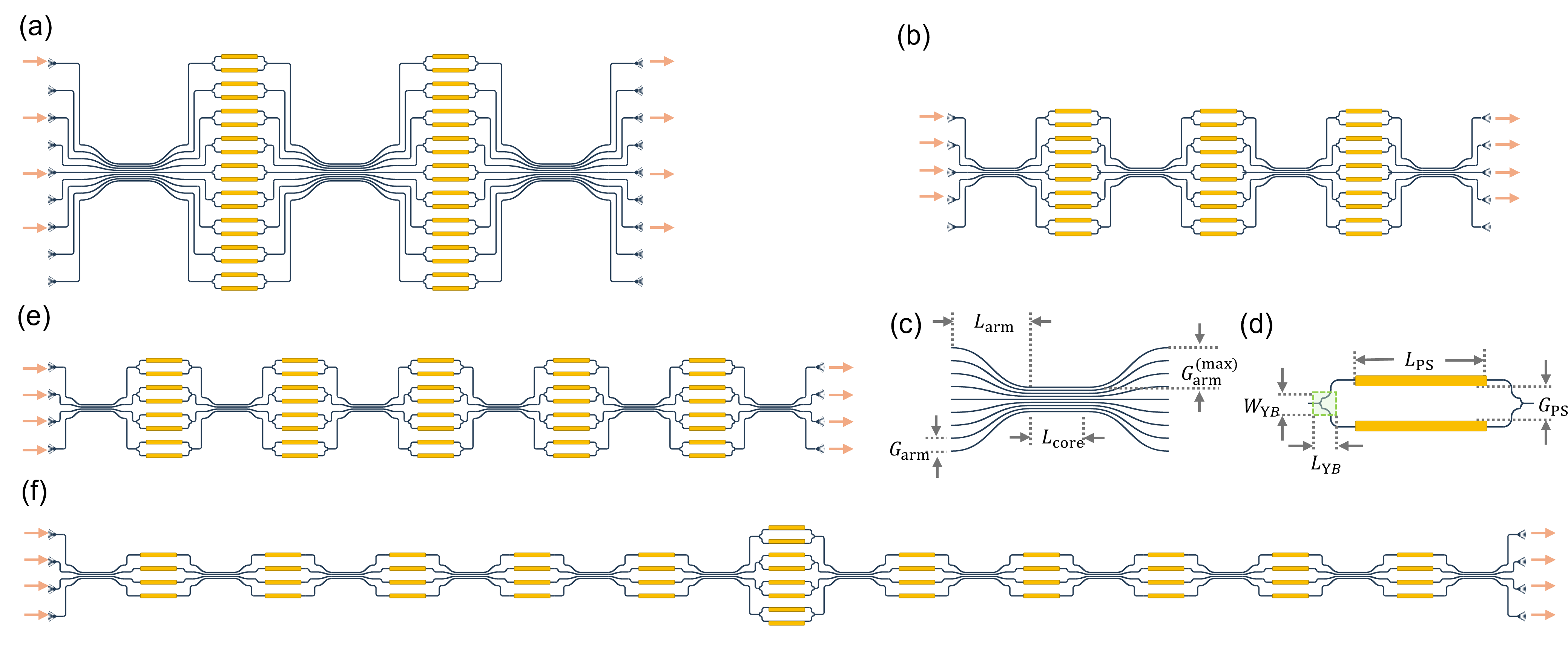}
    \caption{Architecture comparison and scalability. Proposed architecture (lossy layer) with $K=9$ ports and $M=2$ layers (a), and $K=5$ with $M=3$ layers (c), used to reshape a general-purpose $4\times 4$ matrix. Relevant quantities characterizing the waveguide array (c) and MZI (d). (e) Complex-valued matrix implementation reported in~\cite{markowitz2024learning}. (f) Complex-valued matrix implementation through SVD through two unitaries. Figs.~4(a)-(b),(e)-(d) are scaled with the same aspect ratio and relevant design dimensions.}
    \label{fig4}
\end{figure*}

The architecture scalability is assessed by defining some design rules used as the building blocks for an arbitrary number of total ($K$) and effective ($N$) ports. Figs.~\ref{fig4}(a)-(b) depict a sketch for an embedded $4\times 4$ matrices in a $9$-port device with $M=2$ layers and a $5$-port device with $M=2$ layers, respectively, estimated on our previous numerical analysis. The two main components that define the total length of the device are the waveguide array and the amplitude and phase modulation layers. From Figs.~\ref{fig4}(a)-(b), it is clear that the total horizontal length per waveguide array is $L_{\textnormal{WGA}}=L_{\textnormal{C}}+2L_{\textnormal{arm}}+2L_{\textnormal{bend}}$, where $L_{\textnormal{C}}$ is the coupling length,$L_{\textnormal{arm}}$ the waveguide arm length, and $L_{\textnormal{bend}}$ the connecting arm bend. The waveguide arms are built using cubic B\'ezier curves with a length-width ratio $\ell$, where $\ell\geq 3$ typically produces low loss and mode-cross talk. Furthermore, the uppermost arm in the waveguide array possesses the largest curvature in the structure, the gap of which depends on $N$ through the relation $G_{\textnormal{arm}}^{(\textnormal{max})}=\left( \frac{N-1}{2} \right) \left( G_{\textnormal{arm}}-G_{\textnormal{C}} \right) \approx \left( \frac{N-1}{2} \right) G_{\textnormal{arm}}$, where the approximation holds for B\'ezier arm gaps ($G_{\textnormal{arm}}$) much larger than the waveguide coupling core gap ($G_{\textnormal{C}}$). See Fig.~\ref{fig4}(c). The connecting bends couple the waveguide array with the rest of the architecture and are composed of two L-bends, each comprising a circular bend of radius $R$. For symmetry, contiguous connecting bends are kept at the same distance $G_{\textnormal{arm}}$, rendering a total length $L_{\textnormal{bends}}\approx\lfloor \frac{N}{2} \rfloor (G_{\textnormal{arm}})+2R$. 

The tunable layer comprises MZIs (Fig.~\ref{fig4}(d)), composed of two 3-dB Y-branches and two parallel heaters. The utilization of 3 dB Y-branch splitters as bifurcation devices is well-established, which can be straightforwardly manufactured in an effective area of 14.8 µm $\times$ 7 µm and are compatible with conventional single-mode waveguides measuring 500 nm × 220 nm~\cite{zhang2013compact}. In turn, metal heaters are among the most common thermo-optical solutions for phase tuning and device prototyping~\cite{harris2014efficient,liu2022thermo}. The length ($L_{\textnormal{PS}}$) and gaps ($G_{\textnormal{PS}}$) between continuous metal heaters are the relevant quantities that impact the overall design length (Fig.~\ref{fig4}(c)). Proper metal-heater gaps shall be considered to mitigate thermal cross-talk effects, rendering $G_{\textnormal{PS}}\gg W_{\textnormal{YB}}$ and $G_{\textnormal{PS}}\gg W_{\textnormal{PS}}$, with $W_{\textnormal{PS}}$ the width of the metal heater. Under these considerations, the total length for a single MZI becomes $L_{\textnormal{MZI}}=L_{\textnormal{PS}}+2L_{\textnormal{YB}}+4R$ and $W_{\textnormal{MZI}}=G_{\textnormal{PS}}+W_{\textnormal{Heater}}\approx G_{\textnormal{PS}}$. 

The proposed design is compared to other general-purpose solutions, such as the interlaced design reported in~\cite{markowitz2024learning} and the SVD solutions~\cite{miller2013self}. The first one (Fig.~\ref{fig4}(e)) relies on $N+2$ interlacing layers of waveguide arrays intertwined by $N+1$ layers of amplitude and phase modulators, implemented through MZIs layers to optimize the device size. The second design (Fig.~\ref{fig4}(f)) utilizes two universal unitary circuits sandwiching a single layer of MZIs, rendering an all-optical representation of the SVD~\cite{miller2013self}. These two alternative solutions can be represented using the same design rules used for our design, making them suitable for comparison. The sketches in Fig.\ref{fig4} and the design rules show that the proposed design scales vertically as $O(N^2)$. The horizontal scalability strongly depends on the nature of the waveguide separation in the coupling region, which ultimately dictates the length of the waveguide array cores $L_{\textnormal{core}}$. The $J_{x}$ lattice is particularly useful for this task, from which an estimation of $L_{\textnormal{core}}\equiv L_{\textnormal{core}}(N)$ can be assessed. Following the well-known properties of the coupling parameters $\kappa_{n}$ of the $J_{x}$ lattice~\cite{Wei16,zelaya2024integrated}, the coupling $\kappa_{max}:=\kappa_{\lfloor N/2\rfloor}$ takes the largest value. Thus, by fixing $\kappa_{max}$ for different $N$, one can straightforwardly show that $L_{\textnormal{core}}=L_{0}N$, with $L_{0}$ a reference length fixed for $N=2$ and used to scale $L_{\textnormal{core}}$ up to any device size $N$. %See Supplementary Information SI for details. 

\begin{figure}[t]
    \centering
    \includegraphics[width=0.95\linewidth]{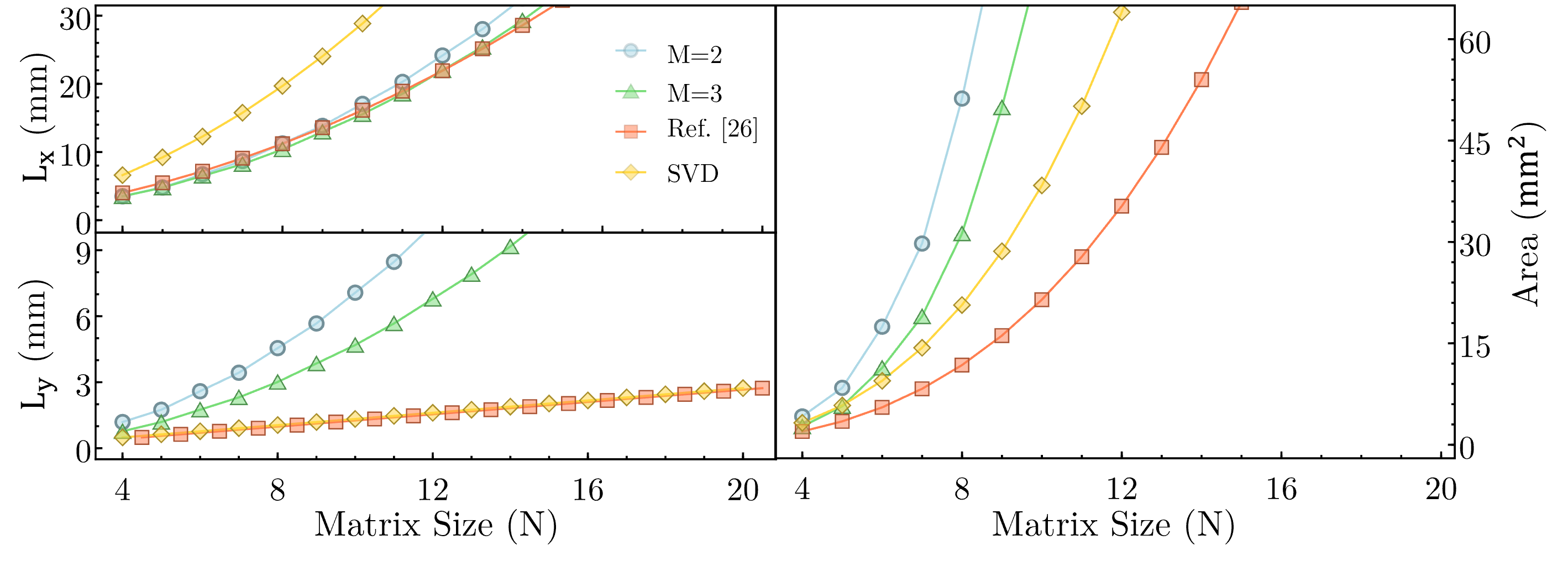}
    \caption{Different architectures scalability shown in terms of the total horizontal $L_{x}$ and vertical $L_{y}$ lengths illustrated in the upper-left and lower-left panels, respectively. The total area is plotted in the right panel.}
    \label{fig5}
\end{figure}

From the previous considerations, the overall size of the three architectures can be established as a function of the total number of ports $N$, which provides a common ground to compare their scales. %See Tab.~SI in Supplementary Information SI for the specifics. 
The total horizontal $(L_{x})$ and vertical $(L_{y})$ lengths and the total device area are illustrated in Figs.~\ref{fig5}(a)-(c). For $M=2$, the vertical length of the proposed architecture surpasses that of its counterparts at the expense of significantly reducing the horizontal length. Nevertheless, the total area still exceeds that of the architecture reported in~\cite{markowitz2024learning}, but it stays below that of the SVD approach. These results suggest that lower-dimensional complex-valued matrices can be universally programmed by embedding them into larger-dimensional matrices through as few as two or three layers of phase shifters.\\

\noindent
\textbf{Funding}. This project is supported by the U.S. Air Force Office of Scientific Research (AFOSR) Young Investigator Program (YIP) Award FA9550-22-1-0189.\\

\noindent
\textbf{Disclosures}. The authors declare no conflicts of interest.

\bibliography{bibfile}

% Note that this extra page will not count against page length
\bibliographyfullrefs{bibfile}

\end{document}